\documentclass[11pt]{article}
\usepackage{feynmf}
\usepackage{amsmath}  % Better maths support
\def\coth{\rm coth}
\def\tanh{\rm tanh}
\def\half{{1\over 2}}

\def\Eins{{\mathchoice {\rm 1\mskip-4mu l} {\rm 1\mskip-4mu l}
{\rm 1\mskip-4.5mu l} {\rm 1\mskip-5mu l}}}
\def\Z{{\mathchoice {\hbox{$\sf\textstyle Z\kern-0.4em Z$}}
{\hbox{$\sf\textstyle Z\kern-0.4em Z$}}
{\hbox{$\sf\scriptstyle Z\kern-0.3em Z$}}
{\hbox{$\sf\scriptscriptstyle Z\kern-0.2em Z$}}}}

\def\square{\kern1pt\vbox{\hrule height 1.2pt\hbox{\vrule width 1.2pt
   \hskip 3pt\vbox{\vskip 6pt}\hskip 3pt\vrule width 0.6pt}
   \hrule height 0.6pt}\kern1pt}
      \def\boxop{{\raise-.25ex\hbox{\square}}}
\def\mn{{\mu\nu}}

\def\tr{{\rm tr}\,}

\def\e{\,{\rm e}}
\def\partder#1#2{{\partial #1\over\partial #2}}

\newcommand{\be}{\begin{equation}}
\newcommand{\ee}{\end{equation}\noindent}
\newcommand{\bear}{\begin{eqnarray}}
\newcommand{\ear}{\end{eqnarray}\noindent}
\newcommand{\benn}{\begin{enumerate}}
\newcommand{\enn}{\end{enumerate}}

\def\slash#1{#1\!\!\!\raise.15ex\hbox {/}}
\newcommand{\slD}{\,\raise.15ex\hbox{$/$}\kern-.27em\hbox{$\!\!\!D$}}
\newcommand{\slpartial}{\raise.15ex\hbox{$/$}\kern-.57em\hbox{$\partial$}}
\def\4piTD{{(4\pi T)}^{-{D\over 2}}}
\def\4piT4{{(4\pi T)}^{-2}}

\def\Tintm4{{\dps\int_{0}^{\infty}}{dT\over T}\,e^{-m^2T}
    {(4\pi T)}^{-2}}
\def\Tintm{{\dps\int_{0}^{\infty}}{dT\over T}\,e^{-m^2T}}

\newcommand{\slG}{{{\dot G}\!\!\!\! \raise.15ex\hbox {/}}}

\def\GBd12{{\dot G}_{B12}}

\newcommand{\no}{\noindent}
\def\non{\nonumber}
\def\dps{\displaystyle}

\begin{document}
%\input feynman

%\special{papersize=8.26in,11.69in}
%\textwidth15.0cm
%\textheight22.0cm
%\baselineskip1.0cm
%\setlength{\topmargin}{-1cm}
%\addtolength{\textheight}{1cm}
%\oddsidemargin+1.2cm
%\evensidemargin-1.2cm

\renewcommand{\thefootnote}{\protect\arabic{footnote}}
\pagestyle{empty}
%------------------------------------------------------

\setcounter{page}{1}
\setcounter{footnote}{0}

\medskip

\begin{center}
{\Large\bf Euler-Heisenberg lagrangians and asymptotic}

\vspace{3pt}

{\bf \Large  analysis in 1+1 QED, part 1: Two-loop}
\vskip1.3cm

{\large I. Huet$^{a}$, D.G.C. McKeon$^{b,c}$, 
C. Schubert$^{a}$}
\\[1.5ex]

\begin{itemize}
\item [$^a$]
{\it 
%   \footnote{}
Instituto de F\'{\i}sica y Matem\'aticas
\\
Universidad Michoacana de San Nicol\'as de Hidalgo\\
Edificio C-3, Apdo. Postal 2-82
\\
C.P. 58040, Morelia, Michoac\'an, M\'exico\\
}
\item [$^b$]
{\it
Department of Applied Mathematics
\\
University of Western Ontario
\\
London, Ontario, N6A 1B7 Canada
}
\item [$^c$]
{\it 
Department of Mathematics and Computer Science\\
Algoma University, Sault Ste. Marie, ON P6A 2G4 Canada
}
\end{itemize}

E-mail: idrish@ifm.umich.mx, dgmckeo2@uwo.ca, schubert@ifm.umich.mx

\end{center}
%\centerline{\today}
\vspace{1cm}
 {\large \bf Abstract:}
%\begin{quotation}
%
We continue an effort to obtain information on the
QED perturbation series at high loop orders, and particularly on the
issue of large cancellations inside gauge invariant classes of
graphs, using the example of the $l$ - loop $N$ - photon amplitudes 
in the limit of large photons numbers and low photon energies.
As was previously shown, high-order information on these
amplitudes can be obtained from a nonperturbative formula, 
due to Affleck et al., 
for the imaginary part of the QED effective lagrangian in a constant
field. The procedure uses Borel analysis and leads, 
under some plausible assumptions, to a number of
nontrivial predictions already at the three-loop level. Their direct verification
would require a calculation of this `Euler-Heisenberg lagrangian' at three-loops,
which seems presently out of reach. Motivated by previous work by Dunne
and Krasnansky on Euler-Heisenberg lagrangians in various dimensions,
in the present work we initiate a new line of attack on this problem
by deriving and proving the analogous predictions in the simpler setting
of 1+1 dimensional QED. 
In the first part of this series, we obtain a generalization of the formula of Affleck et al. 
to this case,  and show that, for both Scalar and Spinor QED, it correctly
predicts the leading asymptotic behaviour of the weak field expansion coefficients
of the two loop Euler-Heisenberg lagrangians. 
%\end{quotation}
\vfill\eject
\pagestyle{plain}
\setcounter{page}{1}
\setcounter{footnote}{0}

\section{Introduction}
\renewcommand{\theequation}{1.\arabic{equation}}
\setcounter{equation}{0}
\bigskip

QED is the oldest and prototypical quantum field theory, and much effort has gone
into calculating low order terms in its perturbation series. However, little is still known about its
high-order behaviour, in fact less than for some other field theories 
\cite{legzin,stevensonprd,stevensonnpb,fischer,dunneminneapolis}. This is due, on one hand, to
the absence of (spacetime) instantons in the abelian case, on the other hand
to large-scale cancellations between Feynman diagrams which are familiar
to all practicioners in the field, but whose origin and effect on the large order
behaviour of perturbation theory is still not well-understood. This type of cancellations
has recently attracted wider attention since it is now increasingly believed that their study may throw light on the 
origin of similar cancellations observed in gravity and supergravity theories
(see, e.g., \cite{babjva,badhen} and refs. therein). 

In their pioneering calculation of the $g-2$ factor of the electron to sixth
order in 1974, Cvitanovic and Kinoshita \cite{cvikin74} found a coefficient which was much
smaller numerically than had been expected by a naive estimate based on
the number of Feynman diagrams involved. A detailed analysis revealed 
extensive cancellations inside gauge invariant classes of diagrams. This led
Cvitanovic \cite{cvitanovic1977} to conjecture that, at least in the quenched
approximation (i.e. excluding diagrams involving virtual fermions) these cancellations
would be important enough numerically to render this series convergent
for the $g-2$ factor (see \cite{cvitanovicweb} for an amusing account of the genesis
of this conjecture). 
Although nowadays there exist many good arguments against convergence of
the QED perturbation series, all of them are based
on the presence of an unlimited number of virtual fermions, so that Cvitanovic's conjecture
for the quenched contribution
is still open today. Moreover, should it  hold true for the case of the $g-2$ factor,
it is natural to assume that it extends to arbitrary QED amplitudes, such as the
quenched photon S-matrix (for photon amplitudes, we call `quenched' the contributions
involving only one scalar/fermion loop). 
Beyond the quenched approximation, it would suggest
that the QED perturbation series should be rearranged as a series in the number
of fermion loops, rather than in the coupling constant. 

The present work continues an effort \cite{dunschSD1,dunschSD2, mascvi, colima} 
to study the multiloop behaviour of the QED $N$ photon amplitudes using the
QED effective lagrangian, and in particular to prove or disprove Cvitanovic's
conjecture for these amplitudes. 
Let us start with recalling the representation obtained by Heisenberg and Euler \cite{eulhei}
for the one-loop QED effective Lagrangian in a constant field,

\bear
{\cal L}^{(1)}_{\rm spin}(F) &=& - \frac{1}{8\pi^2}
\int_0^{\infty}{dT\over T^3}
\,\e^{-m^2T} 
\biggl[
{(eaT)(ebT)\over {\rm tanh} (eaT){\rm tan} (ebT)} 
%\nonumber\\&&\hspace{110pt}
- {1\over 3}(a^2-b^2)T^2 -1
\biggr]
\nonumber\\
\label{eulhei}
\ear
%(the superscript refers to the loop order).
Here $T$ is the proper-time of the loop particle and $a,b$  are defined by 
 $a^2-b^2 = B^2-E^2,\quad  ab = {\bf E}\cdot {\bf B}$.
 The Euler-Heisenberg Lagrangian (`EHL' in the following) contains the information on the
one-loop $N$ photon amplitudes in the limit where all photon energies are small
compared to the electron mass (see, e.g., \cite{itzzub,dunnerev}).   
After expanding the EHL in powers of the field invariants, the `weak field expansion', it is straightforward to
obtain the explicit form of the $N$ photon amplitudes in this `EH limit' from 
the terms in this expansion involving $N$ powers of the field. 
Using the spinor helicity formalism, the result of this procedure can be
expressed quite concisely \cite{mascvi}:

\bear
\Gamma_{\rm spin}^{(EH)}
[\varepsilon_1^+;\ldots ;\varepsilon_K^+;\varepsilon_{K+1}^-;\ldots ;\varepsilon_N^-]
&=&
-{m^4\over 8\pi^2}
\Bigl({2ie\over m^2}\Bigr)^N(N-3)!
\non\\&&\hspace{-90pt}\times
\sum_{k=0}^K\sum_{l=0}^{N-K}
(-1)^{N-K-l}
{B_{k+l}B_{N-k-l}
\over
k!l!(K-k)!(N-K-l)!}
\chi_K^+\chi_{N-K}^- 
\, 
\non\\
\label{resspin}
\ear
Here the superscripts $\pm$ refer to circular polarizations, and the $B_k$ are
Bernoulli numbers. The invariants $\chi_K^{\pm}$ are written, in standard spinor helicity
notation (see, e.g., \cite{dixon}),

\bear
\chi_K^+ &=&
{({\frac{K}{2}})!
\over 2^{K\over 2}}
\Bigl\lbrace
[12]^2[34]^2\cdots [(K-1)K]^2 + {\rm \,\, all \,\, permutations}
\Bigr\rbrace
\non\\
\chi_{N-K}^- &=&
{({\frac{N-K}{2}})!
\over 2^{N-K\over 2}}
\Bigl\lbrace
\langle (K+1)(K+2)\rangle^2\langle (K+3)(K+4)\rangle^2\cdots
\langle (N-1)N\rangle^2 \non\\
&& \hspace{50pt} + {\rm \,\, all \,\, perm.}
\Bigr\rbrace
\non\\
\label{defchiKL+-}
\ear 
The formula analogous to (\ref{eulhei}) for Scalar QED is

\bear
{\cal L}^{(1)}_{\rm scal}(F)&=&  {1\over 16\pi^2}
\int_0^{\infty}{dT\over T^3}
\,\e^{-m^2T} 
\biggl[
{(eaT)(ebT)\over \sinh(eaT)\sin(ebT)} 
%\nonumber\\&&\hspace{110pt}
+{1\over 6}(a^2-b^2)T^2 -1
\biggr]
\non\\
\label{eulheiscal}
\ear
This formula was obtained by Weisskopf \cite{weisskopf}, 
but for simplicity will be called ``scalar EHL" in the following. 

Except for the purely magnetic case,
the effective Lagrangians (\ref{eulhei}),(\ref{eulheiscal}) have an imaginary part.
For the purely electric case, 
Schwinger \cite{schwinger} found the following representation of the
imaginary parts in terms of infinite series of `Schwinger exponentials',

\begin{eqnarray}
{\rm Im}{\cal L}_{\rm scal}^{(1)}(E) 
&=&
-\frac{(eE)^2}{16\pi^3}
\, \sum_{k=1}^\infty \frac{(-1)^{k}}{k^2}
\,\exp\left[-\pi k \frac{m^2}{eE}\right]
\label{schwingerscal}\\
{\rm Im} {\cal L}_{\rm spin}^{(1)}(E) &=&  \frac{(eE)^2}{8\pi^3}
\, \sum_{k=1}^\infty \frac{1}{k^2}
\,\exp\left[-\pi k \frac{m^2}{eE}\right] 
\label{schwingerspin}
\end{eqnarray}
These formulas imply that any constant electric field 
will lead to a certain probability for electron-positron pair creation from
vacuum. The inverse exponential dependence on the field
suggests to think of this as a tunneling process in which virtual pairs
draw enough energy from the field to turn real, as had been proposed by Sauter
as early as 1931 \cite{sauter}.
%In the following we will be interested only in the
%weak field limit $eE/m^2 \ll 1$, which allows us to truncate the 
%series in (\ref{schwingerscal}),(\ref{schwingerspin})
%to the then dominant first ``Schwinger exponential''.

At the one-loop level, obtaining the imaginary part of the EHL from its real part
is a simple application of Cauchy's theorem. For our
multiloop purposes, however, it will be important that the imaginary part can,
using Borel summation,
also be obtained from the asymptotic behaviour of the coefficients of the
weak-field expansion. Referring to \cite{dunneminneapolis,dunsch1} 
for the details of this procedure,
let us just cite the following main result:
Assume that a function $f(g)$ has an asymptotic series expansion 

\bear
f(g) \sim \sum_{n=0}^{\infty} c_n g^n
\label{expf}
\ear
where the expansion coefficients $c_n$ have the leading-order large $n$ behaviour

\bear
c_n \sim \rho^n \Gamma(\mu n + \nu)
\label{cnlargen}
\ear
with some real constants $\rho >0$,  $\mu > 0 $ and $\nu$. Then the series
(\ref{expf}) has convergence radius zero, and is not even Borel-summable. Nevertheless, one can, 
by a dispersion relation applied to the formal Borel integral,
show that the leading imaginary contribution for small $g$ is given by 

\begin{eqnarray}
{\rm Im} f(g)\sim\frac{\pi}{\mu}\left(\frac{1}{\rho g} \right)^{\nu/\mu}
\exp\left[-\left(\frac{1}{\rho g}\right)^{1/\mu}\right]
\label{genimag}
\end{eqnarray}
These formulas can be applied to the weak field expansions of the purely electric EHL's (\ref{eulhei}), (\ref{eulheiscal}),
setting $g=(eE/m^2)^2$. The expansion coefficients $c_n$ involve essentially only the Bernoulli numbers \cite{dunnerev,dunsch1},
and using the asymptotic properties of those numbers it is easy to see that they obey (\ref{cnlargen})
with $\rho = 1/\pi^2$, $\mu = 2$, and $\nu = -2$, independently of spin. Eq. (\ref{genimag})
then reproduces the $k=1$ terms in (\ref{schwingerscal}) and (\ref{schwingerspin}).
One can then go on and iterate this procedure to successively construct all the
Schwinger exponentials \cite{dunsch1}. However, in the
following we will generally be concerned only with the leading term.  

Considerable work has gone into generalizing the EHL's  to the two loop level,
that is, taking a single photon exchange in the loop into account.
For a general constant field, various equivalent integral representations
have been found for these two-loop 
EHL's ${\cal L}_{\rm spin}^{(2)}(F)$ \cite{ritusspin,ginzburg,ditreu,rescsc,frss,korsch,review}
and ${\cal L}_{\rm scal}^{(2)}(F)$ \cite{ritusscal,ginzburg,rescsc,frss,korsch,review}, 
all at the two-parameter integral level.
So far these integrals have withstood attempts at analytical evaluation.
However, they have been used to compute the two-loop weak field expansion coefficients
to high orders \cite{rescsc,frss,dunsch1}, and to derive closed-form expressions for these coefficients in
the purely electric/magnetic case \cite{dhrs}.  

The Schwinger representations (\ref{schwingerscal}),(\ref{schwingerspin}) also
generalize to the two-loop level \cite{lebrit,ginzburg}, although now with a certain prefactor function
$K_k(eE/m^2)$ in front of the $k$th Schwinger exponential; those functions
are known explicitly only at leading orders in the weak-field limit.  

However, it turns out that the electric or magnetic backgrounds are not the simplest ones
in this context. Computationally, the most favorable case is the one of a (euclidean)
self-dual (`SD') field, defined by $F=\tilde F$, which has the consequence that 

\bear
F^2 = -f^2\Eins
\label{Fsquared}
\ear
For a real value of the parameter $f$, the SD effective Lagrangian has properties similar to the magnetic EHL,
for imaginary $f$ similar to the electric one. 
In this SD case, even at two loops it is possible to obtain explicit formulas
for the EHL's \cite{dunschSD1,dunschSD2}:
 
\bear
{\cal L}_{\rm scal}^{(2)(SD)}(\kappa)
&=&
\alpha \,{m^4\over (4\pi)^3}\frac{1}{\kappa^2}\left[
{3\over 2}\xi^2 (\kappa)
-\xi'(\kappa)\right]
\label{L24Dscal}\\
{\cal L}_{\rm spin}^{(2)(SD)}(\kappa)
&=&
-2\alpha \,{m^4\over (4\pi)^3}\frac{1}{\kappa^2}\left[
3\xi^2 (\kappa)
-\xi'(\kappa)\right]
\label{L24Dspin}
\ear
Here we have defined the convenient dimensionless parameter
\begin{eqnarray}
\kappa\equiv \frac{m^2}{2ef}
\label{kappa}
\end{eqnarray}
%where $f^2=\fourth F_{\mu\nu}F^{\mu\nu}$, 
as well as the function
\bear
\xi(x)\equiv -x\Bigl(\psi(x)-\ln(x)+{1\over 2x}\Bigr)
\label{defxi}
\ear
where $\psi$ is the digamma function $\psi(x)=\Gamma^\prime(x)/\Gamma(x)$. 
Thus in the self-dual case the study of the weak-field expansions and the
construction of the imaginary parts involve only some well-known properties
of the digamma function, making it possible to do everything much more
completely and explicitly than for the electric or magnetic case \cite{dunschSD2}.
In particular, for this case it has been verified that certain complications related to
the non-uniqueness of $f$, which could invalidate
the relation between (\ref{cnlargen}) and (\ref{genimag}),  such as the appearance of
additional poles or cuts in the complex $g$ plane, do not occur in QED, or at least not yet at the
two-loop level.

Moreover, although a self-dual field cannot be realized in Minkowski
space, the corresponding EHL still contains information on the $N$ -- photon amplitudes, 
namely on their ``all $+$'' component in the helicity decomposition (\ref{resspin}) \cite{dufish}. 

Beyond the two-loop level, to the best of our knowledge the only result on
EHL's in the literature is the following formula, proposed in 1982 by
Affleck, Alvarez, and Manton \cite{afalma} as an all-loop generalization of the leading Schwinger exponential 
for scalar QED (\ref{schwingerscal}):

\bear
{\rm Im}{\cal L}^{({\rm all-loop})}_{\rm scal}(E)
\,\,
&{\stackrel{E\to 0}{\sim}}&
\,\,
 \frac{(eE)^2}{16\pi^3}
\,{\rm exp}\Bigl[ -\pi  \frac{m^2}{eE}+\alpha\pi \Bigr] 
\label{ImLallloop}
\ear
This formula (called `AAM formula' in the following) is highly remarkable for various reasons.
Despite of its simplicity it is a true all-loop result; the rhs 
receives contributions from an infinite set of 
Feynman diagrams of arbitrary loop order,
as sketched in fig. \ref{aamdiagrams}.

\setlength{\unitlength}{1mm}

%\begin{figure}
\begin{figure}[htbp]
%\begin{center}

\bigskip

\begin{tabular}[c]{|c|c|c|c|c|}
\hline
 & \multicolumn{4}{|c|}{Number of external legs} \\ \hline
%\multicolumn{4}{|c|}{ Rubro: equipo de c\'omputo}\\ \hline
Number of loops & 4 & 6 & 8 & $\cdots$  \\ \hline
 1 & \begin{fmffile}{diagram1}
    \begin{fmfgraph}(30,18)
        \fmfleft{i1,i2}
        \fmfright{o1,o2}
        \fmf{photon}{i1,v1}
        \fmf{photon}{i2,v3}
        \fmf{photon}{v4,o2}
        \fmf{photon}{v2,o1}
        \fmf{plain,right=0.37}{v1,v2}
        \fmf{plain,right=0.37}{v4,v3}
        \fmf{plain,left=0.3}{v1,v3}
        \fmf{plain,right=0.3}{v2,v4}
    \end{fmfgraph}
\end{fmffile} & \begin{fmffile}{diagram2}
    \begin{fmfgraph}(30,18)
        \fmfleft{i1,i2}
        \fmfright{o1,o2}
        \fmftop{u}
        \fmfbottom{b}
        \fmf{plain,right=0.3}{v2,v3,v4}
        \fmf{plain,right=0.3}{v6,v7,v8}
        \fmf{plain,left=0.37}{v2,v8}
        \fmf{plain,left=0.37}{v6,v4}
        \fmf{photon}{i1,v6}
        \fmf{photon}{i2,v4}
        \fmf{photon}{o1,v8}
        \fmf{photon}{o2,v2}
        \fmf{photon}{u,v3}
        \fmf{photon}{b,v7}
        \fmf{phantom,tension=1}{v4,v8}
        \fmf{phantom,tension=1}{v2,v6}
        \fmf{phantom,tension=0.5}{v3,v7}
    \end{fmfgraph}
\end{fmffile} & \begin{fmffile}{diagram3}
    \begin{fmfgraph}(30,18)
        \fmfleft{i1,i2,i3}
        \fmfright{o1,o2,o3}
        \fmftop{u}
        \fmfbottom{b}
        \fmf{plain,right=0.3}{v2,v3,v4}
        \fmf{plain,right=0.3}{v6,v7,v8}
        \fmf{plain,left=.2}{v2,v1,v8}
        \fmf{plain,left=.2}{v6,v5,v4}
        \fmf{photon}{i1,v6}
        \fmf{photon}{i3,v4}
        \fmf{photon}{o1,v8}
        \fmf{photon}{o3,v2}
        \fmf{photon}{u,v3}
        \fmf{photon}{b,v7}
        \fmf{photon}{v1,o2}
        \fmf{photon}{v5,i2}
        \fmf{phantom,tension=1}{v4,v8}
        \fmf{phantom,tension=1}{v2,v6}
        \fmf{phantom,tension=0.4}{v3,v7}
        \fmf{phantom,tension=1.1}{v1,v5}
    \end{fmfgraph}
\end{fmffile} & $\cdots$ \\ \hline
2 & \begin{fmffile}{diagram4}
    \begin{fmfgraph}(30,18)
        \fmfleft{i1,i2}
        \fmfright{o1,o2}
        \fmftop{u}
        \fmfbottom{b}
        \fmf{plain,right=0.3}{v2,v3,v4}
        \fmf{plain,right=0.3}{v6,v7,v8}
        \fmf{plain,left=0.37}{v2,v8}
        \fmf{plain,left=0.37}{v6,v4}
        \fmf{photon}{i1,v6}
        \fmf{photon}{i2,v4}
        \fmf{photon}{o1,v8}
        \fmf{photon}{o2,v2}
        \fmf{photon,tension=0}{v3,v7}
        \fmf{phantom,tension=1}{v4,v8}
        \fmf{phantom,tension=1}{v2,v6}
        \fmf{phantom,tension=0.5}{v3,v7}
        \fmf{phantom,tension=1}{u,v3}
        \fmf{phantom,tension=1}{b,v7}   
     \end{fmfgraph}
\end{fmffile} & 
\begin{fmffile}{diagram5}
    \begin{fmfgraph}(30,18)
        \fmfleft{i1,i2,i3}
        \fmfright{o1,o2,o3}
        \fmftop{u}
        \fmfbottom{b}
        \fmf{plain,right=0.3}{v2,v3,v4}
        \fmf{plain,right=0.3}{v6,v7,v8}
        \fmf{plain,left=.2}{v2,v1,v8}
        \fmf{plain,left=.2}{v6,v5,v4}
        \fmf{photon}{i1,v6}
        \fmf{photon}{i3,v4}
        \fmf{photon}{o1,v8}
        \fmf{photon}{o3,v2}
        \fmf{photon}{u,v3}
        \fmf{photon}{b,v7}
        \fmf{photon}{v1,v5}
        \fmf{phantom,tension=1}{v1,o2}
        \fmf{phantom,tension=1}{v5,i2}
        \fmf{phantom,tension=1}{v4,v8}
        \fmf{phantom,tension=1}{v2,v6}
        \fmf{phantom,tension=0.4}{v3,v7}
    \end{fmfgraph}
\end{fmffile} & $\cdots$ & $\cdots$ \\ \hline
3 & \begin{fmffile}{diagram6}
    \begin{fmfgraph}(30,18)
        \fmfleft{i1,i2,i3}
        \fmfright{o1,o2,o3}
        \fmftop{u}
        \fmfbottom{b}
        \fmf{plain,right=0.3}{v2,v3,v4}
        \fmf{plain,right=0.3}{v6,v7,v8}
        \fmf{plain,left=.2}{v2,v1,v8}
        \fmf{plain,left=.2}{v6,v5,v4}
        \fmf{photon}{i1,v6}
        \fmf{photon}{i3,v4}
        \fmf{photon}{o1,v8}
        \fmf{photon}{o3,v2}
        \fmf{photon}{v3,v5}
        \fmf{photon}{v1,v7}
        \fmf{phantom}{u,v3}
        \fmf{phantom}{b,v7}
        \fmf{phantom}{v1,o2}
        \fmf{phantom}{v5,i2}
        \fmf{phantom,tension=1}{v4,v8}
        \fmf{phantom,tension=1}{v2,v6}
        \fmf{phantom,tension=0.3}{v3,v7}
        \fmf{phantom,tension=1.1}{v1,v5}
        %%corrective phantoms
        \fmf{phantom,tension=0.3}{u,v3}
        \fmf{phantom,tension=0.3}{o3,v3}
        \fmf{phantom,tension=0.3}{i2,v5}
        \fmf{phantom,tension=0.3}{i1,v5}
        \fmf{phantom,tension=0.3}{v7,b}
        \fmf{phantom,tension=0.3}{v7,i1}
         \fmf{phantom,tension=0.3}{v1,o2}
        \fmf{phantom,tension=0.3}{v1,o3}
    \end{fmfgraph}
\end{fmffile} & $\cdots$ & $\ddots$ & $\vdots$ \\ \hline
$\vdots$ & $ \vdots$ & $\ddots$ & $\ddots$ & $\vdots$ \\ \hline
\end{tabular}

\bigskip

\caption{{Diagrams contributing to ${\rm Im}{\cal L}^{({\rm all-loop})}_{\rm scal}(E)$
in the weak-field limit.  }}
\label{aamdiagrams}
%\end{center}
\end{figure}

\bigskip
\bigskip
\bigskip
\bigskip

Moreover, the mass appearing in (\ref{ImLallloop}) is argued to be still the physical renormalized
mass, which means that the above figure should strictly 
speaking include also the mass renormalization counter diagrams 
which appear in EHL calculations starting from two loops.  

The derivation given in \cite{afalma} is very
simple, if formal. Based on a stationary path approximation of
Feynman's worldline path integral representation \cite{feynman50} of
${\cal L}_{\rm scal}(E)$, it actually uses only
a one-loop semiclassical trajectory, and arguments that this
trajectory remains valid in the presence of virtual photon
insertions. This also implies that non-quenched diagrams do not
contribute in the limit (\ref{ImLallloop}), which is why we have
shown only the quenched ones in fig. \ref{aamdiagrams}.

Although the derivation of (\ref{ImLallloop}) in \cite{afalma} cannot be considered
 rigorous,  an independent heuristic derivation of (\ref{ImLallloop}), as well as extension to
 the spinor QED case (with the same factor of $\e^{\alpha\pi}$) was given by Lebedev
 and Ritus \cite{lebrit} through the consideration of higher-order corrections
 to the pair creation energy in the vacuum tunneling picture. 
 At the two-loop level,  (\ref{ImLallloop}) and its spinor QED extension state that
 
 \bear
 {\rm Im}{\cal L}^{({\rm 2})}_{\rm scal,spin}(E)
\,\,
&{\stackrel{E\to 0}{\sim}}&
\,\,
\alpha\pi\, {\rm Im}{\cal L}^{({\rm 1})}_{\rm scal,spin}(E)
\label{AAM2loop}
\ear
which has been verified by  a direct calculation of the EHL \cite{dunsch1} (for the
spinor QED case), and also been extended to the case of a 
self-dual field \cite{dunschSD1,dunschSD2}.
%For the electric EHL case this could be done only numerically, but for the
%self-dual case even analytically.
 
Now, writing the AAM formula (\ref{ImLallloop}) as
(hereafter, unless indicated otherwise, we refer to the scalar and spinor QED
cases simultaneously)

\bear
{\rm Im}{\cal L}^{({\rm all-loop})}(E)\,\,
=
\,\,\sum_{l=1}^{\infty}{\rm Im}{\cal L}^{(l)}(E)
\,\,
&{\stackrel{E\to 0}{\sim}}&
\,\,
{\rm Im}{\cal L}^{(1)}(E)\,\,{\rm e}^{\alpha\pi}
\label{LalltoL1}
\ear
%where ${\cal L}^{(l)}_{\rm scal}(E)$ denotes the $l$-loop generalization of the
%electric EHL.
it states that an all-loop summation has produced the convergent
factor $\e^{\alpha\pi}$, clearly an observation similar in vein to Cvitanovic's.
Moreover, at a formal level it is not 
difficult to transfer this loop summation factor from 
${\rm Im}{\cal L}(E)$ to the QED  photon amplitudes \cite{dunschSD2,colima}.
Consider the weak field expansion of the $l$-loop contribution to the electric
EHL:

\bear
{\cal L}^{(l)}(E) &=& \sum_{n=2}^{\infty} c^{(l)}(n) \Bigl(\frac{eE}{m^2}\Bigr)^{2n}
\label{wfe}
\ear
Using the Borel dispersion relations discussed above, (\ref{ImLallloop}) can be shown \cite{dunsch1,dunschSD2} 
to imply that, at any loop order $l$, the weak field expansion coefficients 
have the same leading asymptotic growth as we found above for the one-loop case,
that is

\bear
&& c^{(l)}(n)\quad {\stackrel{n\to \infty}{\sim}} \quad c^{(l)}_{\infty}\, \pi^{-2n}\Gamma(2n - 2)
\label{asymp}
\ear
where the constant $c^{(l)}_{\infty}$  relates directly to the prefactor of the
corresponding leading Schwinger exponential in the weak field limit:

\bear
{\rm Im}{\cal L}^{(l)}(E)\,\, &{\stackrel{E\to 0}{\sim}} &\,\,  c^{(l)}_{\infty}\,\e^{-\pi \frac{m^2}{eE}}
\label{borel}
\ear
We would now like to convert (\ref{borel}) into an equation for photon amplitudes,
using the above-mentioned correspondence between the weak field expansion coefficients
and the EH limit of the photon amplitudes.  Here we encounter the problem that this relation
cannot be applied to a purely electric field; we will therefore now switch to the SD case,
and, on the strength of the  two-loop results of \cite{dunschSD1,dunschSD2},
assume that (\ref{LalltoL1}) holds for the SD case unchanged. 

Now, the SD EHL relates to the `all +'  $N$ - photon amplitudes, and for those  
the whole kinematic structure in the EH limit can, independently of the loop order,
be absorbed into the invariant $\chi_N^+$ defined in (\ref{defchiKL+-}) \cite{mascvi}.
One can thus eliminate this kinematic factor by dividing the $l$ - loop amplitude by the one-loop one.
Expanding (\ref{ImLallloop}) in $\alpha$ and combining it with (\ref{borel}) and (\ref{asymp}) 
one then  arrives at a formula for this ratio of amplitudes in the limit of large photon number \cite{dunschSD2},

\bear
{\rm lim}_{N\to\infty}
\frac
{\Gamma^{(l)}[k_1,\varepsilon_1^+;\ldots ;k_N,\varepsilon_N^+]}
{\Gamma^{(1)} [k_1,\varepsilon_1^+;\ldots ;k_N,\varepsilon_N^+]}
&=&
{{\rm lim}_{n\to\infty}} {c^{(l)}(n)\over c^{(1)}(n)} 
=
\frac{({\alpha\pi})^{l-1}}{(l-1)!}
\non\\
\label{ratio}
\ear
If we could now sum both sides over $l$ and interchange the sum and
limit, we could reconstruct the $\e^{\alpha\pi}$ factor, and conclude
that the perturbation series for the $N$ - photon amplitudes,
if only for sufficiently large $N$, in the low energy limit and 
for this polarization component, has infinite
convergence radius. 
But this is too good to be true, since so far we have nowhere made a
distinction between quenched and unquenched contributions to
the photon amplitudes, and convergence of the whole
perturbation series can certainly be excluded.  

However, as was noted in \cite{colima} this distinction comes in
naturally if one takes into account that, as mentioned above,
in the path integral derivation of
(\ref{ImLallloop}) in \cite{afalma} the rhs comes entirely
from the quenched sector; all non-quenched contributions are suppressed
in the weak field limit. And since (switching back to the usual Feynman diagram
picture) the importance of non-quenched diagrams is growing with
increasing loop order, it is natural to assume that their inclusion will slow
down the convergence towards the asymptotic limit with increasing $l$,
sufficiently to invalidate the above naive interchange of limits. 
On the other hand, there is no obvious reason to expect such a slowing
down of convergence inside the quenched sector, which led to the
prediction \cite{colima} that Cvitanovic's ``quenched convergence''
will indeed be found to hold true for the photon amplitudes.

%\section{Importance of checks at the three-loop level:}
To further corroborate this prediction, one should now
calculate the EHL's and their weak field expansion
coefficients $c^{(3)}(n)$ at the three-loop level. This would already
allow one to perform some quite nontrivial checks on the
above chain of reasoning, namely:

\bigskip
\bigskip

\benn

\item
The asymptotic relation 

\bear
{{\rm lim}_{n\to\infty}} {c^{(3)}_{\rm scal,spin}(n)\over c^{(1)}_{\rm scal,spin}(n)} 
=
\frac{(\alpha\pi)^2}{2} 
\label{ratio31}
\ear
and its independence of spin.

\item
The absence of a slow-down of the convergence of the ratio (\ref{ratio31})
as compared to the corresponding two-loop to one-loop ratio.

\item
The asymptotic subdominance of the non-quenched part of the EHL, which
first appears at three loops.

\enn

However, a calculation of any three-loop EHL,
be it in Scalar or Spinor QED, for an electric or self-dual field,
presently still poses an enormous computational challenge.

Now, it is well-known that the structure of the EHL's and associated
Schwinger exponentials at one-loop is essentially independent of the
space-time dimension \cite{blviwi,gavgit,gagigo,lin}.
In particular, the Schwinger formulas for 1+1 dimensional QED are different from
 (\ref{schwingerscal}), (\ref{schwingerspin}) only by normalization factors:

\begin{eqnarray}
{\rm Im} {\cal L}_{\rm spin}^{(1)(2D)}(E) &=&  \frac{eE}{4\pi}
\, \sum_{k=1}^\infty \frac{1}{k^2}
\,\exp\left[-\pi k \frac{m^2}{eE}\right] 
\label{schwingerspin2D}\\
{\rm Im}{\cal L}_{\rm scal}^{(1)(2D)}(E) 
&=&
-\frac{eE}{4\pi}
\, \sum_{k=1}^\infty \frac{(-1)^{k}}{k^2}
\,\exp\left[-\pi k \frac{m^2}{eE}\right]
\label{schwingerscal2D}
\end{eqnarray}
More surprisingly,  in 2006 Dunne and Krasnansky \cite{dunkra,krasnansky} found
that, for the scalar QED case, even at the two-loop level
there persists a strong similarity between the 2D EHL
and the 4D self-dual EHL. For the former, they find

\bear
{\cal L}_{\rm scal}^{(2)(2D)}(\kappa)
&=&
-\frac{m^2}{2\pi} \frac{\tilde\alpha}{32}
\left[
\xi^2_{2D} 
-4\kappa \xi_{2D}'\right] 
\label{L22Dscal}
\ear
where $\tilde\alpha := 2\frac{e^2}{\pi m^2}$ is our definition of the fine structure constant
in two dimensions, and

\bear
\xi_{2D}(\kappa)&:=& -\Bigl(\psi(\kappa+\half)-\ln (\kappa)\Bigr)
=  \psi(\kappa)-2\psi(2\kappa) + \ln (4\kappa)
\non\\
\label{defxi2D}
\ear
Here the parameter $f$ is defined by (see app. \ref{conventions} for
our 2D QED conventions)

\bear
F &=& \left(\begin{array}{cc}0 & f \\-f & 0 \end{array}\right)
\label{defF}
\ear
(note that this definition is consistent with (\ref{Fsquared})).  
The formal similarity between (\ref{L22Dscal}) and the 4D SD EHL (\ref{L24Dscal})
has led us to consider 2D QED as a toy model for studying the above asymptotic predictions. 
This will make sense only, of course, if the AAM formula (\ref{ImLallloop2D}) can be generalized to
the 2D case. As we will show below, this generalization is

\bear
{\rm Im}{\cal L}^{({\rm all-loop})(2D)}_{\rm scal}(E)
\,\,
&{\stackrel{E\to 0}{\sim}}&
\,\,
 \frac{eE}{4\pi}
 \e^{-\frac{m^2\pi}{eE} + \tilde\alpha \pi^2  \kappa^2}
\label{ImLallloop2D}
\ear
As in the 4D case, we shall assume that spin does not play a role in this limit, so that
(\ref{ImLallloop2D}) holds for spinor QED unchanged (let us also mention here that Loskutov et al. \cite{lolysk} 
have found an all-loop exponentiation formula similar to (\ref{ImLallloop2D}) for the strong-field asymptotics
of the electron mass operator in a two-dimensional approximation of four-dimensional QED).  
%(note that in the 2D case
%the $k=1$ terms in (\ref{schwingerscal2D}) and (\ref{schwingerspin2D}) agree including the
%normalization factors). 

In the same way as in the 4D case, one can then use Borel analysis to derive from
(\ref{ImLallloop2D}) a formula for the limits of ratios of $l$ - loop to one - loop
coefficients:

\bear
{{\rm lim}_{n\to\infty}} {c^{(l)}_{2D}(n)\over c^{(1)}_{2D}(n+l-1)} 
&=& {(\tilde\alpha\pi^2)^{l-1}\over (l-1)!} 
\label{AAM2Dcoeff}
\ear
Here our definition of the expansion coefficients in 2D is

\bear
{\cal L}^{(l)(2D)}(\kappa) &=& \frac{m^2}{2\pi} \sum_{n=1}^{\infty}(-1)^{l-1}c_{2D}^{(l)}(n) (i\kappa)^{-2n}
\label{defcn}
\ear
Note that, due to the factor of $\kappa^2$ appearing in the exponent on the rhs of
(\ref{ImLallloop2D}), the  2D formula (\ref{AAM2Dcoeff}) involves also a shift in the
argument of $c_{2D}^{(1)}(n)$; the leading asymptotic growth
of the coefficients increases with the loop order. This actually simplifies matters,
since it implies that, for our asymptotic purposes, there is no need to perform
mass renormalization: Mass renormalization contributions to an EHL at the $l$ - loop level
are of the form 

\bear
\delta m^{(l_1)} \delta m^{(l_2)} \cdots \delta m^{(l_j)}  
\frac{\partial^j}{\partial m^j} {\cal L}^{(l')}
\label{massrengen}
\ear
with some $l' < l$ and $l-l'=\sum_{i=1}^j l_i$, where $m^{(l_i)}$
denotes an $l_i$ - loop mass counterterm. 
As we will show below, $c_{2D}^{(1)}(n)$ grows like  $\Gamma(2n-1)$, 
and (\ref{AAM2Dcoeff}) then predicts that
$c_{2D}^{(l)}(n)$ grows like $\Gamma(2n+2l-3)$. It is then easily seen that the
expansion coefficients of a term of the structure (\ref{massrengen}) at $l$ loops can  
grow at most like $\sim \Gamma(2n+2l-4)$, and thus are subdominant.

This is very different from the 4D case. Here the leading asymptotic growth of the
expansion coefficients of the unrenormalized electric EHL's ${\cal L}_{\rm scal,spin}^{(l)(4D)}$ is
like $\Gamma(2n+l-3)$ ($\Gamma(2n+l-2)$ for the SD case), that is, the argument of the $\Gamma$ - function jumps only
by one per loop order \cite{colima}. Thus mass renormalization terms {\sl do} contribute at
the leading asymptotic level, and are even crucial to make the AAM formula work:
 At two loops, it has been shown \cite{dunsch1,dunschSD2} 
that, precisely when the mass is taken to be the physical renormalized mass, the leading asymptotic terms cancel between
the main contribution and the one from mass renormalization, which reduces the asymptotic growth at two loops to make it
the same as at one-loop, as is implicit in the AAM formula (\ref{ImLallloop}). And for this formula to hold, these
cancellations between main and counterterms must not only persist, but become increasingly extensive at higher loop
orders. Thus  the AAM formula already predicts that the asymptotic behaviour of QED must depend crucially on
whether mass renormalization is done physically or just generically.

In section \ref{aam} we will first retrace the original derivation of the AAM formula and its generalization to spinor QED \cite{wlinst1},
and then present our generalization to the 2D QED case. In sections \ref{1loop} and \ref{2loop} we compute the one- and two- loop
EHL's in 2D spinor QED. In section \ref{sectionwfe} we use these results, together with the formulas previously obtained
in \cite{dunkra,krasnansky} for
scalar QED, to verify that the asymptotic predictions derived from our generalized AAM
formula hold at two loop order in both theories. 
Section \ref{conc} gives a summary.      
The second part of this series \cite{part2} will be devoted to the calculation of the three-loop EHL's in scalar and spinor 2D QED,
and to the verification of the 2D analogue of the three-loop predictions discussed above.

\no
Some preliminary results of this work have been presented in \cite{qfext}.

\section{Predictions of the AAM formalism}
\label{aam}
\renewcommand{\theequation}{2.\arabic{equation}}
\setcounter{equation}{0}
\bigskip
In this chapter, we will first retrace the derivation of the AAM formula
(\ref{ImLallloop}), and then use their method
to obtain also the generalization of this formula to the 2D case, eq. (\ref{ImLallloop2D}).

The seminal work of those authors  concerned scalar QED in four dimensions, therefore let us 
start with this case, beginning at the one-loop level. 
The (euclidean) one-loop effective action for scalar QED can be written in the
following
way \cite{feynman50}:

\bear
\Gamma_{\rm scal} [A] &=&
\int_0^{\infty}\frac{dT}{T}\, \e^{-m^2T}
\int_{x(T)=x(0)} {\mathcal D}x
\, \e^{-S[x(\tau)]} \non\\
S[x(\tau)] &=&
\int_0^Td\tau
\left(\frac{\dot x^2}{4} +i e A\cdot \dot x \right)
\label{PI}
\ear
Here $m$ is the mass of the scalar particle,
and the functional integral $\int {\mathcal D}x$ is over all closed spacetime
paths $x^\mu(\tau)$ which are periodic in the proper-time parameter $\tau$,
with period $T$.
Rescaling $\tau = Tu$, the effective action may be expressed as
\bear
&&\Gamma_{\rm scal} [A] =\\
&&\, \, \, \int_0^{\infty}\frac{dT}{T}\, \e^{-m^2T}
\int_{x(1)=x(0)} {\mathcal D}x
\, {\rm exp}\left[-\left(\frac{1}{4T}\int_0^1du \,
\dot x^2 +i e\int_0^1du \, A\cdot \dot x
\right)\right] \non
\label{PIscale}
\ear
%where the functional integral $\int {\mathcal D}x$ is now over closed
%paths $x^\mu(u)$ with period $1$.
After this rescaling we can perform the proper-time integral using the
method of steepest descent.
The $T$ integral has a stationary point at
\bear
T_0 = \frac{1}{2m}\sqrt{\int_0^1 du \, \dot x^2}
\label{defT0}
\ear
leading to
\bear
{\rm Im}\, \Gamma_{\rm scal} =
{1\over m}\sqrt{\pi\over T_0}
\,{\rm Im} \int {\cal D}x \,
\e^{-\Bigl(m\sqrt{\int \dot x^2}
+ie\int_0^1 du A\cdot \dot x
\Bigr)}
\label{elTint}
\ear
Here we have implicitly used the large mass approximation
\bear
m\sqrt{\int_0^1 du\,\dot{x}^2}\gg 1 \quad 
\label{large}
\ear

The functional integral remaining in the effective action expression
(\ref{elTint}) may be approximated by a further, functional,
stationary phase approximation. The new worldline ``action'',
\bear
S_{\rm eff} = m\sqrt{\int_0^1 du\, \dot x^2} + i e \int_0^1duA\cdot \dot x
\label{defS}
\ear
is stationary if the path $x_\alpha(u)$ satisfies
\bear
m{\ddot x_{\mu}\over \sqrt{\int_0^1 du\, \dot x^2}} &=& i e F_{\mn}\dot x_{\nu}
\label{statcond}
\ear
A periodic solution $x_\mu(u)$ to (\ref{statcond}) is called a ``worldline
instanton''.
Contracting (\ref{statcond}) with $\dot{x}_\mu$ shows that for such an
instanton
\bear
\dot{x}^2={\rm constant}\equiv a^2
\label{c2}
\ear
Generally, the existence of a worldline instanton for a background $A$
leads to an imaginary part in the effective action $\Gamma_{\rm scal}[A]$, and
the leading behavior is
\bear
{\rm Im}\Gamma_{\rm scal}[A]\sim e^{-S_0}\quad 
\label{leading}
\ear
where $S_0$ is the worldline action (\ref{defS}) evaluated on the worldline
instanton.

For a constant electric background of magnitude $E$, pointing in the
$z$ direction, the Euclidean gauge field is
$A_3(x_4) = -iEx_4$. The instanton equation (\ref{statcond}) for this case can
be easily
solved, and the solutions are simply circles in the $z-t$ plane
of radius $\frac{m}{eE}$ \cite{afalma}:

\bear
x_k^3(u)=\frac{m}{eE}\,\cos(2 k \pi u)\quad , \quad
x_k^4(u)=\frac{m}{eE}\,\sin(2 k \pi u)
\label{circle}
\ear
(with $x_{1,2}$ kept constant).
The integer $k\in {\bf Z}^+$
counts the number of times the closed path is traversed, and
the instanton action (\ref{defS}) becomes

\bear
S_0 := S_{\rm eff}[x_k^{\mu}] &=&
%\frac{m}{a}\, \left(\frac{2 n \pi m}{e E}\right)^2 \int_0^1 \cos(2 n \pi u)^2
2k\,\frac{m^2\pi}{eE} - k\,\frac{m^2\pi}{eE} = k\,\frac{m^2\pi}{eE}
\label{action-constant}
\ear
Thus in the large mass approximation (\ref{large})
the contribution of the instanton with winding number $k$  reproduces the
exponent of the $k$th term of Schwinger's formula (\ref{schwingerscal}).

Proceeding to the multiloop case, Feynman's formula (\ref{PI}) can be easily modified
to take into account the effect of multiple exchanges of photons in the scalar loop.
This requires only the addition of the following interaction term $S_i$ to the worldline
action,

\bear
S_i &=&
{e^2\over 8\pi^2}\int_0^Td\tau_1\int_0^Td\tau_2 {\dot x(\tau_1)\cdot\dot x(\tau_2)\over
(x(\tau_1)-x(\tau_2))^2}\nonumber\\
\label{defSi}
\ear
After this addition in (\ref{PI}), $\Gamma_{\rm scal}[A]$ turns into the ``quenched'' effective action, 
corresponding to the set of all Feynman
diagrams with a single scalar loop (but any number of interactions with the external field and internal photon exchanges). 
Now, Affleck et al. \cite{afalma} argue that the instanton solutions
(\ref{circle}) remain stationary even in the presence of this addition, so that its effect in the
large mass limit is only a modification of the stationary action. Evaluation of $S_i$ on 
the leading $k=1$ instanton yields simply

\bear
S_i[x(u)] &=& -\alpha \pi
\label{evalSi}
\ear
Combining (\ref{leading}),(\ref{action-constant}),(\ref{evalSi})
one obtains the large mass approximation
for the quenched multiloop scalar EHL (\ref{ImLallloop}),

\bear
{\rm Im}{\cal L}^{({\rm all-loop})}_{\rm scal}(E)
\,\,
&{\stackrel{E\to 0}{\sim}}&
\,\,
 \frac{(eE)^2}{16\pi^3}
\,{\rm exp}\Bigl[ -\pi  \frac{m^2}{eE}+\alpha\pi \Bigr] 
\nonumber
\ear
Affleck et al. then proceed to show that the contributions to the full effective action
involving more than one scalar loop are suppressed with respect to
the quenched one in this large mass limit, so that 
(\ref{ImLallloop}) actually holds even for the full effective action.
Even more remarkably, they argue that
the asymptotic formula (\ref{ImLallloop}) already takes into
account all the effects of mass renormalization which come into play 
for this effective action starting with the two-loop level.

The approach of Affleck et al. was generalized in \cite{wlinst1} to the case of
spinor QED. The path integral representation 
of the one-loop effective action due to a spin half particle in the loop differs from (\ref{PI}) 
only by a global factor of $-\half$ and the insertion of the following `spin factor' $S[x,A]$ under the path integral \cite{feynman51},
\bear
S[x,A] = \tr_{\Gamma}{\cal P}\,\,\e^{{i\over 2}e\sigma^{\mn}\int_0^Td\tau F_{\mn}(x(\tau))}
\label{defspinfactor}
\ear   
Here $\tr_{\Gamma}$ denotes the Dirac trace and ${\cal P}$ the path ordering operator. 
The presence of this spin factor term in the stationary path approximation
is again taken into account simply by evaluating this term on the stationary path 
(\ref{circle}). The result turns out to be  a simple global factor \cite{wlinst1}, 

\bear
4\cos(\pi n)=4(-1)^n
\label{constantspin}
\ear
Taking the global factor of $-\half$ into account this correctly reproduces
the difference between the one-loop Schwinger formulas for scalar and
spinor QED (\ref{schwingerscal}),(\ref{schwingerspin}). 
Proceeding to spinor QED at the multiloop level, 
%although the internal photon exchange term (\ref{defSi}) can be generalized
%to the spinor case using the ``worldline super formalism'' \cite{review},
here there seems to be no simple way to directly calculate the effect of the spin term 
in the stationary path approximation. 
However, as was discussed already in the introduction
the physical arguments of \cite{lebrit} as well as explicit two-loop calculations
strongly suggest that the AAM formula (\ref{ImLallloop}) holds also for the
spinor QED case (up to a global factor of $1/2$). 
%Schwinger formulas for scalar QED at the one-loop and two-loop levels  
%should, in the large mass limit, differ only by a factor of $\alpha\pi$. As 
%we have seen in () this is confirmed by the explicit two-loop 
%result for the scalar QED case, and the same factor appears
%also in the ratio of the two-loop and one-loop Schwinger formulas
%for spinor QED. Moreover, we have seen that the Lebedev-Ritus interpretation
%of the higher loop contributions to ${\rm Im}\Gamma(E)$ in terms of 
%corrections to the tunneling energies for pair production naturally
%lead to the same exponential factor, independently of whether the
%tunneling particles have spin zero or spin half.   

We will now generalize the AAM formula to the case of QED
in two dimensions. First, we note that the path integral representation (\ref{PI}) for
the scalar QED effective action is independent of dimension;
the only explicit appearance of the 
dimension $D$ is in the normalization of the free path integral, which
is  

\bear
\int_{x(T)=x(0)} {\mathcal D}x 
\, {\rm exp}\left[-\int_0^Td\tau \,
\frac{\dot x^2}{4} \right]
&=& (4\pi T)^{-D/2}
\label{PIfree}
\ear
(with the conventions of \cite{review}). This does not affect any of the
manipulations which we performed at the one-loop level, including the form
of the worldline instanton (\ref{circle}), so that the 
asymptotic estimate (\ref{leading}) remains valid in $D=2$ with the
same formula (\ref{action-constant}) for $S_0$. 
%This is borne out
%also by the fact that the exponents of the Schwinger-formulas () 
%are known to be universal; only the prefactors depend on the spacetime dimension \cite{lin}. 

Things are different at the multiloop level. Here it must be remembered that
the worldline insertion term (\ref{defSi}) actually involves the photon propagator
in $D=4$ and in Feynman gauge. For general $D$ and in a general covariant
gauge this term reads

\begin{eqnarray}
S_i(D) &=& {e^2\over 2}
{1\over {4{\pi}^{{D\over2}}}}
\int_0^T d\tau_a \int_0^T d\tau_b
\Biggl\lbrace
{{1+\alpha}\over 2}\Gamma\Bigl({D\over 2}-1\Bigr)
{{\dot x_a\cdot\dot x_b\over
{\Bigl[{(x_a - x_b)}^2\Bigr]}^{{D\over 2}-1}}\quad
}\nonumber\\
&&+(1-\alpha)\Gamma\Bigl({D\over 2}\Bigr)
{\dot x_a\cdot(x_a-x_b)
(x_a-x_b)\cdot\dot x_b\over
{\Bigl[{(x_a - x_b)}^2\Bigr]}^{D\over 2}\quad
}
\Biggr\rbrace
\label{ccigengauge}
\end{eqnarray}
\noindent
where $\alpha = 1$ corresponds to Feynman gauge. Since the
first term in braces in (\ref{ccigengauge}) becomes singular in
$D=2$, in this case instead of Feynman gauge it is more convenient
to choose the gauge $\alpha=-1$, leading to

\begin{eqnarray}
S_i^{D=2} &=& {e^2\over 4\pi }
\int_0^T d\tau_a \int_0^T d\tau_b
{\dot x_a\cdot(x_a-x_b)
(x_a-x_b)\cdot\dot x_b\over
{(x_a - x_b)}^2\quad
}
\label{Si2D}
\end{eqnarray}
\noindent
The evaluation of $S_i^{D=2}$ on the constant field worldline instanton (\ref{circle}) 
then yields

\bear
S_i^{D=2}[x(u)] &=&  {\pi\over 2} {m^2\over E^2} = 2\pi (i\kappa)^2 {e^2\over m^2}
=  -\tilde\alpha \pi^2  \kappa^2
\label{evalSi2D}
\ear
This brings us to our 2D generalization of the AAM formula, eq. (\ref{ImLallloop2D}).

\section{1 loop Euler-Heisenberg Lagrangian}
\label{1loop}
\renewcommand{\theequation}{3.\arabic{equation}}
\setcounter{equation}{0}
\bigskip
The (renormalized)
one-loop Euler-Heisenberg Lagrangians in 2D scalar and spinor QED are, in the standard
proper-time representation (see, e.g., \cite{blviwi}),

\bear
{\cal L}_{\rm scal}^{(1)}(f) &=& {ef\over 4\pi}
\int_0^{\infty}{dz\over z}\e^{-2\kappa z}
\Bigl(\frac{1}{\sinh(z)} - {1\over z}\Bigr) 
\label{L1scal}
\\
{\cal L}_{\rm spin}^{(1)}(f)&=& -{ef\over 4\pi}
\int_0^{\infty}{dz\over z}\e^{-2\kappa z}
\Bigl(\coth(z) - {1\over z}\Bigr) 
\label{L1spin}
\ear
It is convenient to observe that the scalar EHL can be written in terms of the spinor EHL as \cite{huet}

\bear
{\cal L}_{\rm scal}^{(1)}(f) = {\cal L}_{\rm spin}^{(1)}(f)- 2{\cal L}_{\rm spin}^{(1)}(f/2)
\label{reladolfo}
\ear
The integrals (\ref{L1scal}),(\ref{L1spin}) have various equivalent closed-form representations; 
the most suitable one for our purposes is

\bear
{\cal L}_{\rm spin}^{(1)}(\kappa ) &=& -{m^2\over 4\pi} {1\over\kappa}
\Bigl[{\rm ln}\Gamma(\kappa) - \kappa(\ln \kappa -1) +
\half \ln \bigl({\kappa\over 2\pi}\bigr)\Bigr]
\label{L1expl}
\ear

\section{2 loop Euler-Heisenberg Lagrangian}
\label{2loop}
\renewcommand{\theequation}{4.\arabic{equation}}
\setcounter{equation}{0}
\bigskip
We now calculate the two-loop Euler-Heisenberg Lagrangian.
In momentum space, it is given by

\bear
{\cal L}_{\rm spin}^{(2)} (f) &=& {e^2\over 2}
\int {d^2p\over (2\pi)^2}
\int {d^2p'\over (2\pi)^2}
\,\tr \Bigl[g(p)\sigma_{\mu}g(p')\sigma_{\mu}\Bigr]
{1\over (p-p')^2}
\label{L2}
\ear
After plugging in (\ref{gp}), we note that due to (\ref{simp})
the terms involving $\sigma\cdot p$ 
do not contribute under the Dirac trace,
which thus becomes simply

\bear
\tr \bigl[ \e^{\sigma^3 z} \sigma_{\mu} \e^{\sigma^3 z'} \sigma_{\mu}\bigr]
&=&
\tr \bigl[ \e^{\sigma^3 z} \e^{-\sigma^3 z'}\sigma_{\mu} \sigma_{\mu}\bigr]
=
4 \cosh(z-z')
 \label{simptr}
 \ear
This leaves us with

\bear
{\cal L}_{\rm spin}^{(2)} (f) &=& 2e^2m^2
\int_0^{\infty}dT\int_0^{\infty}dT'
\e^{-m^2(T+T')}
{\cosh(z-z')\over \cosh z \cosh z'} 
\non\\
&&\times
\int {d^2p\over (2\pi)^2}
\int {d^2p'\over (2\pi)^2}
{1\over (p-p')^2}
\e^{-{1\over  ef}(\tanh z\, p^2 + \tanh z'\, p'^2)}
\label{L2p}
\ear
We exponentiate the photon propagator,

\bear
{1\over (p-p')^2}
&=&
\int_0^{\infty} d\lambda \e^{-\lambda (p-p')^2}
\label{expphot}
\ear
The momentum integrals can now be done, yielding

\bear
\int d^2p
\int d^2p'
 \e^{-\lambda (p-p')^2-{1\over  ef}(\tanh z\, p^2 + \tanh z'\, p'^2)}
&=&
{\pi^2 (ef)^2 \over \tanh z\tanh z' + \lambda\, ef (\tanh z + \tanh z')}
\non\\
\label{evalp}
\ear
The $\lambda$ - integral has a logarithmic infrared divergence 
which we regulate with a cutoff $\lambda_0$,
$\int_0^{\infty}d\lambda \to \int_0^{\lambda_0} d\lambda$.
Using

\bear
{1\over \cosh z \cosh z' (\tanh z + \tanh z')} &=& 
{1\over \sinh (z+z')}
\ear
the result becomes, in the limit $\lambda_0ef \gg 1$,

\bear
{\cal L}_{\rm spin}^{(2)} (f) &=& {e^3f m^2\over 8\pi^2}
\int_0^{\infty}\!\!dT\int_0^{\infty}\!\!dT'\e^{-m^2(T+T')}
{\cosh(z-z')\over\sinh(z+z')}
\Bigl[ \ln (ef\lambda_0) + \ln \Bigl({\sinh(z+z')\over\sinh z \sinh z'}\Bigr)\Bigr] \non\\
\label{L2u}
\ear
For the calculation of the integrals, it is convenient to change to the ``worldline''
variables \cite{review},

\bear
T' &=& u\tilde T \non\\
T &=& (1-u) \tilde T \non\\
\int_0^{\infty} dT \int_0^{\infty}dT' &=& \int_0^{\infty}d\tilde T \tilde T \int_0^1 du \non\\
\label{cov}
\ear
There are two different $u$ integrals, both elementary:

\bear
\int_0^1 du \cosh\bigl[(1-2u)Z\bigr]&=& {\sinh(Z)\over Z} \non\\
\int_0^1 du \cosh\bigl[(1-2u)Z\bigr] \ln \sinh (uZ) &=& 
-\half \cosh Z + {\sinh Z\over Z}\bigl(\ln\sinh Z -\half\bigr)\non\\
\label{integrals}
\ear
where $Z \equiv ef\tilde T$.
After some rearrangements, the result can be written as

\bear
{\cal L}_{\rm spin}^{(2)}(f) &=& {m^2e^2\over 8\pi^2} {1\over ef}
\int_0^{\infty}dZ \e^{-2\kappa Z} \Bigl[Z(\coth Z -{1\over Z}) - \ln\sinh Z + \ln Z \Bigr]\non\\
&& + {e^2\over 8\pi^2} \Bigl[ \ln (\lambda_0 m^2) + 2 + \gamma \Bigr] 
\label{L2Z}
\ear
($\gamma$ is the Euler-Mascheroni constant).
The remaining integrals can be easily reduced to the standard integral \cite{dunschbernoulli}

\bear
\int_0^{\infty} dZ \e^{-2\kappa Z} (\coth Z -{1\over Z}) &=& \frac{\xi(\kappa)}{\kappa}
\label{intcosh}
\ear 
%where
%
%\bear
%\tilde\psi (x) &\equiv& \psi(x) - \ln x + {1\over 2x}
%\label{defpsitilde}
%\ear
Putting the pieces together, our final result for the (unrenormalized) 2-loop spinor EHL in 2D is

\bear
{\cal L}_{\rm spin}^{(2)}(\kappa) &=& {m^2\over 4\pi}\frac{\tilde\alpha}{4}
\Bigl[-\xi'(\kappa)
%\tilde\psi(\kappa) + \kappa \tilde\psi'(\kappa)
+\ln(\lambda_0 m^2) + \gamma + 2 \Bigr]
\label{L2fin}
\ear
%Finally, renormalization removes the field-independent part of the Lagrangian, leading to
%
%\bear
%{\cal L}^{(2)}_{\rm ren}(f) &=& {e^2\over 8\pi^2}
%\Bigl[ \tilde\psi(\kappa) + \kappa \tilde\psi'(\kappa)\Bigr]
%\label{L2finren}
%\ear
%Note that this also removes the dependence on the infrared cutoff.

Thus, contrary to the scalar QED case, the two-loop 2D spinor QED EHL turns out to
be significantly simpler than the corresponding self -- dual 4D one (\ref{L24Dspin}).
In fact, it has the curious property that, up to the vacuum part, one can even write it in terms of derivatives
of the one-loop EHL (\ref{L1expl}):

\bear
{\cal L}_{\rm spin}^{(2)}(\kappa) &=& - \frac{\tilde\alpha}{4}\Bigl(m^2\partder{}{m^2}\Bigr)^2 {\cal L}_{\rm spin}^{(1)}(\kappa) 
\label{L2byL1}
\ear
Note that, as explained in the introduction, there is no need for us to perform the
mass renormalization of ${\cal L}_{\rm spin}^{(2)}$. 

\section{Weak field expansions}
\label{sectionwfe}
\renewcommand{\theequation}{5.\arabic{equation}}
\setcounter{equation}{0}

Finally, we will now work out the weak field expansions of the one and two-loop $2D$ EHL's,
for both the scalar and spinor cases, and verify the asymptotic prediction (\ref{AAM2Dcoeff})
derived from the AAM formalism.  
  
Starting with the spinor case,  for working out 
the one-loop EHL (\ref{L1expl}) we need the large $x$ expansion of
$\ln \Gamma (x)$, which is

\bear
\ln\Gamma (x) \sim x(\ln x -1 )-\half\ln \bigl(\frac{x}{2\pi} \bigr)
+\sum_{n=1}^{\infty} {B_{2n}\over 2n(2n-1)} x^{-(2n-1)}
\label{lnGlargex}
\ear
This yields (the subscript `2D' will be omitted in this chapter)

\bear
c_{\rm spin}^{(1)}(n) &=& (-1)^{n+1} \frac{B_{2n}}{4n(2n-1)}
\label{c1loopspin}
\ear
($n\geq 1$).
At the two-loop level, we get from (\ref{lnGlargex}) that

\bear
%\tilde\psi (x) + x\tilde\psi'(x) 
-\xi'(x)
&\sim&
\sum_{n=1}^{\infty}
\frac{2n-1}{2n}B_{2n}x^{-2n}
\label{psilargex}
\ear
Thus (\ref{L2fin}) gives (omitting the cutoff-dependent vacuum energy term)

\bear
c_{\rm spin}^{(2)}(n) &=& (-1)^{n+1} \frac{\tilde\alpha}{8}\frac{2n-1}{2n}B_{2n}
\label{c2loopspin}
\ear
($n\geq 1$).
Using Euler's formula 

\bear
B_{2n}&=&(-1)^{n+1}2(2\pi)^{-2n}(2n)!\zeta(2n)
\label{euler}
\ear
and $\lim_{n\to\infty}\zeta(n)=1$ we find

\bear
c_{\rm spin}^{(1)}(n) &\sim& \frac{\Gamma(2n-1)}{(2\pi)^{2n}} \non\\
c_{\rm spin}^{(2)}(n) &\sim& \frac{\tilde{\alpha}}{4}\frac{\Gamma(2n+1)}{(2\pi)^{2n}} \non\\
\label{c12spin}
\ear
so that

\bear
\lim_{n\to\infty}  {c_{\rm spin}^{(2)}(n)\over c_{\rm spin}^{(1)}(n+1)} 
&=&
\tilde\alpha \pi^2
\label{ratio21}
\ear
in agreement with (\ref{AAM2Dcoeff}).

Proceeding to the scalar QED case, at one-loop we get, from 
(\ref{reladolfo}) and (\ref{c1loopspin}),

\bear
c_{\rm scal}^{(1)}(n) &=& (-1)^{n+1} (1-2^{1-2n})\frac{B_{2n}}{4n(2n-1)}
\label{c1loopscal}
\ear
($n\geq 1$). Comparing with (\ref{c1loopspin}), we see that 
$c_{\rm scal}^{(1)}(n) \sim c_{\rm spin}^{(1)}(n)$ for large $n$.
%(which relates by Borel analysis to the fact that the leading Schwinger
%exponentials in (\ref{schwingerscal2D}) and  (\ref{schwingerspin2D}) coincide). 
At two loops, we need the expansion of the function $\xi_{2D}(\kappa)$, 
defined in (\ref{defxi2D}), which is 

\bear
 \xi_{2D}(\kappa) = \sum_{n=1}^{\infty}\frac{\bar B_{2n}}{2n}\kappa^{-2n}
\label{expandxi2D}
\ear
where we have defined $\bar B_n := (2^{1-n}-1)B_n$. 
The expansion of (\ref{L22Dscal}) then gives

\bear
c_{\rm scal}^{(2)}(n) = \frac{\tilde\alpha}{8} (-1)^n \Bigl( \sum_{m=1}^{n-1}\frac{\bar B_{2m}}{4m}\frac{\bar B_{2n-2m}}{4(n-m)}
+ \bar B_{2n} \Bigr)
\label{c2loopscal}
\ear
($n\geq 2$). The structure of these coefficients is thus very similar to the one of the
expansion coefficients of the two-loop EHL's for a self-dual field
in 4D \cite{dunschSD1,dunschSD2}. In \cite{dunschSD2} also a method was
developed to compute the asymptotic expansion of folded sums
of Bernoulli numbers of the type appearing in the first term in brackets in (\ref{c2loopscal});
applying the same technique to the case at hand one can show that this term
is subdominant in the large $n$ limit. The leading order contribution comes from
the second term, and yields

\bear
c_{\rm scal}^{(2)}(n) &\sim& \frac{\tilde{\alpha}}{4}\frac{\Gamma(2n+1)}{(2\pi)^{2n}} \non\\
\label{c2loopscallim}
\ear
Comparing with (\ref{c2loopspin}), we see that
the two-loop coefficients  $c_{\rm scal}^{(2)} (n)$ and $c_{\rm spin}^{(2)} (n)$
also agree for large $n$. This completes our check of (\ref{ratio21}) for the
scalar case.

\section{Conclusions}
\label{conc}
\renewcommand{\theequation}{6.\arabic{equation}}
\setcounter{equation}{0}

To summarize, in the first part of this series we have, generalizing the
work of Affleck et al., obtained an all-loop formula for the 
imaginary part of the Euler-Heisenberg lagrangian in 1+1
QED, in the limit of a weak field but at arbitrary coupling.
 We have performed the first calculation of the Euler-Heisenberg
lagrangian in 2D spinor QED at two loops, and verified that
the asymptotic behaviour of its weak field expansion coefficients 
agrees with the prediction of this generalized AAM formula. 
We have also performed the same check for the scalar QED case,
using the two-loop result of Dunne and Krasnansky. 
Our findings clearly demonstrate that 2D QED is
sufficiently close to the 4D case to suggest that it may hold
generic information on the asymptotic behaviour of amplitudes
in QED in general, but at the same time holds more promise for
the explicit study of multiloop Euler-Heisenberg lagrangians.
A particularly interesting common aspect of our two-loop 2D QED results
and the  two-loop SD EHL calculations of \cite{dunschSD1,dunschSD2} is, that,
in all cases where an explicit formula has been obtained for a two-loop EHL,
the leading asymptotic growth of its expansion coefficients has turned out to involve 
only the digamma function, and only linearly. We take this as further evidence that
the study of the $N$-photon amplitudes through the QED effective Lagrangian 
may ultimately provide a window to high orders in perturbation theory.

\no
{\bf Acknowledgements:} C. S. thanks G.V. Dunne and H. Gies for discussions. I. H.
and C. S. thank CONACYT for financial support. D.G.C. M. thanks the IFM, UMSNH for
hospitality during part of this work.

%\vfill\eject

\begin{appendix}

\section{Conventions and formulas for Euclidean 1+1 QED}
\label{conventions}
\renewcommand{\theequation}{A.\arabic{equation}}
\setcounter{equation}{0}
\bigskip

\no
{\it Dirac equation:}

\bear
\Bigl(\sigma_{\mu}(\partial_{\mu}-ieA_{\mu}) + m\Bigr) \psi &=& 0
\label{diraceq}
\ear

\no
{\it Free electron propagator:}

\bear
{1\over i\slash{p} + m} &=& {-i\slash{p} +  m\over p^2 +m^2}
\label{freeelprop}
\ear
($\slash{p}=\sigma_{\mu}p_{\mu}$).

\no
{\it Photon propagator in Feynman gauge:}

\bear
{\delta_{\mn}\over k^2}
\label{phoprop}
\ear

\no
{\it Vertex:} 

\bear
ie\sigma_{\mu}
\label{vertex}
\ear

\no
{\it Constant field:}

\bear
F &=& \left(\begin{array}{cc}0 & f \\-f & 0 \end{array}\right)
\label{defFapp}
\ear

\no
{\it Fock-Schwinger gauge:}

\bear
A_{\mu}(x) &=& -\half F_{\mn} x_{\nu}
\label{FSgauge}
\ear

\no
{\it Electron propagator in a constant field in Fock-Schwinger gauge:}

\bear
\bigl[\sigma_{\mu}(\partial_{\mu} -ieA_{\mu})+m \bigr] g(x-x') &=& \delta (x-x')
\label{defgx}
\ear

\bear
g(p) &=& \int_0^{\infty} dT \e^{-T\bigl(m^2+{\tanh z\over z}p^2\bigr)}
{1\over\cosh z} \Bigl(m\e^{\sigma^3 z} - {i\slash{p}\over\cosh z}\Bigr)
\label{gp}\\
g(x) &=& \int {d^2p\over (2\pi)^2} \e^{ip\cdot x}g(p) \label{gx}\\
&=&
{1\over 4\pi} \int_0^{\infty} {dT\over T} \e^{-m^2T}{z\over\sinh z}
\e^{-{z\over \tanh z}{x^2\over 4T}}
\Bigl(m\e^{\sigma^3 z} +{1\over 2T} {z\over \sinh z} \slash{x}\Bigr)
\non
\ear 
($z=efT$).

One of the motivations for considering the Euler-Heisenberg Lagrangian 
in $1+1$ dimensions is that substantive simplifications can be expected for 
higher loop calculations in Feynman gauge due to the fact that 

\bear
\sigma_{\mu}\sigma_{\nu_1}\cdots \sigma_{\nu_{2n+1}}\sigma_{\mu}
&=& 0
\label{simp}
\ear

\end{appendix}

\end{document}